\documentclass[apjl, trackchanges, twocolumns]{emulateapj}

\usepackage{xcolor} 
\usepackage[backref,breaklinks,colorlinks,urlcolor=blue, citecolor={blue!50!black}]{hyperref}
\usepackage{bm}
\usepackage{calc}
\usepackage{amsmath}
\usepackage{afterpage}
\usepackage{appendix}

\hypersetup{colorlinks,linkcolor={red!50!black}, citecolor={blue!50!black},urlcolor={blue!80!black} } 

\shorttitle{Detection of missing baryons in groups with kSZE}
\shortauthors{Lim et al.}

\begin{document}

\title{Detection of missing baryons in galaxy groups 
with kinetic Sunyaev-Zel'dovich effect}

\author{S.H. Lim\altaffilmark{1,2}, 
H.J. Mo\altaffilmark{1},
Huiyuan Wang\altaffilmark{3},
Xiaohu Yang\altaffilmark{4}}

\altaffiltext{1}
{Department of Astronomy, University of Massachusetts, Amherst MA 01003-9305}

\altaffiltext{2}
{Department of Physics and Astronomy, University of British Columbia, 6224 Agricultural Road, Vancouver, BC V6T 1Z1, Canada; \textcolor{blue}{slim@phas.ubc.ca}}

\altaffiltext{3}
{Key Laboratory for Research in Galaxies and Cosmology, Department of Astronomy, 
University of Science and Technology of China, Hefei, Anhui 230026, China; 
School of Astronomy and Space Science, University of 
Science and Technology of China, Hefei 230026, China}

\altaffiltext{4}
{Department of Astronomy, Shanghai Key Laboratory for Particle Physics and Cosmology, Shanghai Jiao Tong University, Shanghai 200240, China; 
IFSA Collaborative Innovation Center, and Tsung-Dao Lee Institute, 
Shanghai Jiao Tong University, Shanghai 200240, China}

\begin{abstract} 
We present the detection of the kinetic Sunyaev-Zel'dovich effect 
(kSZE) signals from groups of galaxies as a function of halo mass down to 
$\log (M_{500}/{\rm M_\odot}) \sim 12.3$, using the {\it Planck} CMB
maps and stacking about $40,000$ galaxy systems with known positions, halo masses, 
and peculiar velocities. The signals from groups of different mass are 
constrained simultaneously to take care of 
projection effects of nearby halos. The total kSZE flux within halos 
estimated implies that the gas fraction in halos is about the universal 
baryon fraction, even in low-mass halos, indicating that the 
`missing baryons' are found. Various tests performed show that our results are 
robust against systematic effects,  such as contamination by infrared/radio 
sources and background variations, beam-size effects and 
contributions from halo exteriors. Combined with the thermal Sunyaev-Zel'dovich 
effect, our results indicate that the `missing baryons' associated with 
galaxy groups are contained in warm-hot media with temperatures between 
$10^5$ and $10^6\,{\rm K}$. 
\end{abstract} 



\section[intro]{INTRODUCTION}
\label{sec_intro}

According to the current scenario of galaxy formation, 
galaxies form and evolve in dark matter halos 
\citep[see][for a review]{mo10}. As a dark matter halo forms in 
the cosmic density field, the cold gas associated with it
falls into its potential well and gets shock-heated, eventually  
forming a hot gaseous halo, with a temperature roughly equal to 
the virial temperature of the halo. However, various processes, 
such as radiative cooling, star formation, and feedback from 
supernovae and active galactic nuclei (AGN), can affect the evolution 
of galaxies and the properties of the gaseous halos, so that the distribution 
of baryons may be very different from that of the dark matter. 
Indeed, observations have shown that both the hot gas fraction and the 
total baryon fraction in present-day galaxy systems are much lower than 
the universal baryon fraction, especially in low-mass systems 
\citep[e.g.][]{david06, gastaldello07, pratt09, sun09}. 
Even for massive clusters of galaxies, the distribution of
the gas is found to be different from that of the dark matter, 
although the total amount of the hot gas is found to be close to the 
universal fraction \citep[e.g.][]{arnaud10,battaglia12}. 
It has been suggested that a significant portion 
of the `missing baryons' may be in the form of diffuse warm-hot 
intergalactic media (WHIM), with temperature in the range 
of $10^5$ - $10^7\,{\rm K}$, within and/or around dark matter 
halos \citep[e.g.][]{cen99, dave99, dave01, smith11}.
Attempts have been made to detect such WHIM
using X-ray observations \citep[e.g.][]{werner08, 
eckert15} and quasar absorption line systems
\citep[e.g.][]{fang10,tumlinson13}. 

The Sunyaev-Zel'dovich effect \citep[SZE;][]{sunyaev72} offers another
promising way to probe the WHIM \citep{hill18, lim18b, degraaff19, tanimura19}. 
As the cosmic microwave background  
(CMB) photons pass through galaxy systems, such as clusters and groups of 
galaxies, they are scattered by the free electrons in these systems. 
The effect produced on the CMB by the thermal motion of electrons 
is referred to as the thermal SZE (tSZE), while that produced by 
the bulk motion of electrons is called the kinetic SZE (kSZE). 
Thus, cross-correlating galaxy systems (clusters and groups 
of galaxies, collectively referred to as galaxy groups hereafter)
and their SZE in the CMB provides an avenue to probe the 
WHIM associated with dark matter halos. 

Great efforts have been made to measure the tSZE  
from observations and to use it to constrain the gas associated with 
galaxy systems. \citet{pcxi} used the {\it Planck} 
multi-frequency CMB temperature maps and dark matter halos identified
based on isolation criteria, to study the tSZE down to a halo mass 
$\sim 4\times 10^{12}{\rm M_\odot}$. Remarkably, they found that the hot gas 
fraction in halos is independent of halo mass, as expected from the
simple self-similar model. Similarly, \citet{greco15} used the locally 
brightest galaxies to represent dark matter halos to extract the tSZE and found 
that their results are consistent with the self-similar model. 
In a recent paper, \citet{lim18a} used a large sample of galaxy groups 
\citep{lim17b} to extract the tSZE associated with galaxy systems from 
the {\it Planck} Compton parameter map \citep{pcxxii}. By stacking 
about half a million galaxy systems, they were able to obtain the 
tSZE as a function of halo mass down to $\log (M_{500}/{\rm M_\odot}) \sim 12$, 
where $M_{500}$ is the halo mass enclosed by a radius in which the 
mean mass density is 500 times the critical density. They found that 
the thermal contents of the gas in low-mass halos are much lower than that 
expected from the cosmic mean baryon fraction and the virial temperature of 
halos, in contrast to the results obtained by \citet{pcxi} and \citet{greco15}. 
Tests by \citet{lim18a} and \citet{hill18} suggest that the 
discrepancy may originate from different treatments of projection 
effects in the measurements. 

Detecting the kSZE signals from CMB observations is not a trivial task.
First, the signals are weak: the kSZE amplitude is 
about two orders of magnitude smaller than the 
primary CMB fluctuation, and typically one order of magnitude 
lower than the tSZE. 
As such, stacking a large number of similar 
galaxy systems is needed to detect the effect. Second, since the kSZE 
is directly proportional to the radial peculiar velocity of the 
galaxy system, and since the peculiar velocities of different 
systems have a symmetric distribution around zero, 
stacking individual systems without using the peculiar
velocity information leads to cancellation rather than 
enhancement of signals. Third, the large beam 
sizes of current CMB experiments require assumptions of the locations 
and gas profiles of the galaxy systems to be stacked, in order to extract 
the kSZE they produce.  Finally, since the observed effects are projected 
on the sky, signals from low-mass systems may be contaminated by projections of 
the more massive systems along the same line-of-sight. 

The detection of kSZE has so far been made only for cluster-size individual 
systems \citep[e.g][]{sayers13, adam17} or from statistical measurements based 
on, e.g., the pair-wise and cross correlation methods
\citep{hand12, HM15, pcxxxvii, hill16, schaan16, soergel16, 
debernardis17}. Using peculiar velocity fields 
reconstructed from galaxy distributions and the aperture 
photometry, \citet{HM15, pcxxxvii, schaan16} found 
that a significant fraction of baryons may be associated with 
the large-scale structure traced by galaxies. A similar conclusion 
was reached by \citet{hill16} by cross-correlating galaxies with CMB maps. 
The signals measured in these investigations are the 
averages over individual galaxies in the galaxy samples 
used, including effects of the gas both confined to galaxy halos 
and unbound over large scales. 
These results, therefore, constrain the total amount of free 
electrons associated with the large-scale structure traced 
by galaxies, but cannot be interpreted directly in terms of 
baryon fractions in halos of different masses. 
Thus, the missing baryon problem on halo scales,  
which has important implications for galaxy formation
in dark matter halos, remains unresolved.  

In this paper, we investigate the kSZE from halos of different mass, 
using galaxy groups and the {\it Planck} temperature maps, 
by extending methods similar to those used in \citet{lim18a} to kSZE.
As described below, our analysis differs from earlier studies in that our
halo-based group catalog with reliable halo mass information
allows us to probe the kSZE and baryon fractions in halos of 
different masses, and that we simultaneously constrain 
signals from different groups so that the line-of-sight 
contamination by projection effects is taken into account. 
In addition, the combination of the kSZE measurements
here with the tSZE measurements obtained in \citet{lim18a}
not only allows us to obtain the total mass, but also  
the effective temperature of the WHIM associated with galaxy systems. 

The structure of the paper is as follows. We describe the observational
data for our analysis in Section~\ref{sec_data}, and our method   
to extract the kSZE in Section~\ref{sec_methods}. 
We present our main results and inferences from combining kSZE with 
tSZE in Section~\ref{sec_results}. Finally, we summarize and conclude in 
Section~\ref{sec_sum}.


\section[data]{OBSERVATIONAL DATA}
\label{sec_data}

\subsection{The \textit{Planck} CMB map} 
\label{ssec_map}

The {\it Planck} observation \citep{tauber10, pci} 
measures the all-sky CMB anisotropy in nine frequency bands from $30$ to $857\,$GHz, 
with angular resolutions ranging from $31\arcmin$ to $5\arcmin$. In our analysis 
for the kSZE, we use the $100$, $143$, and $217\,$GHz channel maps from the {\it 
Planck} 2015 data release\footnote{\url{https://pla.esac.esa.int}}. 
To minimize Galactic contamination, the brightest $40\%$ of the sky is masked 
using the masks provided in the data release.  
We also eliminate known radio and infrared point sources using the 
corresponding masks. From the reduced maps, subtractions are made of 
the tSZE, 
\begin{eqnarray}
\left(\frac{\Delta T}{T_{\rm CMB}}\right)_{\rm tSZ} = 
g(x) y \equiv 
g(x)\frac{\sigma_{\rm T}}{m_{\rm e}c^2} \int{P_{\rm e} {\rm d}l}, 
\end{eqnarray}
where $T_{\rm CMB}=2.7255\,{\rm K}$, $y$ is the Compton parameter, 
$g(x)=x \coth(x/2)-4$ is the conversion factor at a given 
$x\equiv h\nu/(k_{\rm B}T_{\rm CMB}$), $\sigma_{\rm T}$ is the 
Thompson cross-section, $c$ is the speed of light, 
$m_{\rm e}$ is the electron rest-mass, and 
$P_{\rm e}=n_{\rm e}k_{\rm B}T_{\rm e}$ is the electron pressure with 
$n_{\rm e}$ and $T_{\rm e}$ being the free electron density and 
temperature, respectively. The electron pressure is integrated over the 
path length, ${\rm d}l$, along the line-of-sight (LOS).
We adopt the Compton parameter $y$ from the {\it Planck} NILC 
\citep[Needlet Independent Linear Combination;][]{remazeilles11}
all-sky $y$-map \citep{pcxxii}, which is constructed from the full mission 
data set of the {\it Planck}, using a combination of different frequency maps 
to minimize the primary CMB fluctuations and contamination from foreground sources. 
Integrating over the {\it Planck} bands gives 
the conversion factor $g(x)T_{\rm CMB}=-4.031$, $-2.785$, and 
$0.187\,{\rm K}$ for the $100$, $143$, and $217\,$GHz maps, respectively. 
The readers are referred  
to the original papers for more details about the constructions of the $y$-maps. 
Finally, a constant background is subtracted from each of the resulting maps 
to zero the average background over the full-sky. 
As a test, we also applied the same analysis to the {\it Planck} 
MILCA \citep[Modified Internal Linear Combination Algorithm;][]{hurier13} $y$-map, 
which is known to have a different level of dust contamination, and found
no significant changes in our results. 

The power spectra of dust emission from galaxies in 
the {\it Planck} maps at the frequencies relevant to the SZE analysis 
are more than $1,000$ times smaller than the CMB at $l\sim 100$, 
and are subdominant in comparison to the primordial  
CMB for all $l$ \citep{pcxxx}. In addition, the dust emission 
was found to be only weakly correlated with galaxy systems at low-$z$ 
\citep[e.g.][]{vikram17}, 
as is expected from the fact that most of the dust emission  
comes from star-forming galaxies at higher redshift. 
Additional leakages of the dust emission through the subtraction of 
$y$-maps may be present in the maps but are very poorly constrained 
\citep{vikram17}. 
As long as such dust contamination is not correlated with 
the targets used for our analysis (see \S\ref{ssec_group}), 
it will only add noise to the data but will not bias our results, 
and the level of the noise will be much lower than that 
produced by the primary CMB fluctuations. 

Thus, the final maps we use in our analysis still  
contain components other than the kSZE, such as the primary CMB, 
dust emission, instrumental noise and residual contaminants. 
The uncertainties introduced by these contaminating components 
will be taken into account in our analysis, as described 
in \S\ref{sec_methods}.

\subsection{Galaxy groups} 
\label{ssec_group}

To extract the SZ signals associated with galaxy groups requires 
a well-defined group catalog. In our analysis, we use the group
catalog of \citet{yang07}, which is constructed from the 
Sloan Digital Sky Survey Data Release 7 \citep[SDSS DR7;][]{abazajian09}
with the use of the halo-based group finder developed in \citet{yang05}. 
All the groups in the original catalog have accurate estimates of 
halo masses, spatial positions, and peculiar velocities. Halo masses 
of the groups are estimated from abundance-matching based on the ranking of their 
characteristic luminosities. Tests using realistic mock catalogs show 
that the halo masses given by the group finder match well 
the true halo masses, with typical scatter of $0.2-0.3\,{\rm dex}$.  
Following conventions in SZE studies, we define a halo 
by a radius, $R_{500}$, within which the mean density is $500$ times 
the critical density at the redshift of the halo. The corresponding 
halo mass is $M_{500}$. The masses provided in the 
group catalog, $M_{200}$, are converted to $M_{500}$ by
assuming the NFW profile \citep{navarro97} and a model for the 
halo concentration parameter \citep{neto07}. These properties of 
individual groups are used in a filter to extract kSZE from galaxy groups 
over a large range of masses. They are also used to interpret the kSZE in 
terms of the total amount of ionized gas associated with these groups. 
We adopt the radial peculiar velocities, $v_r$, reconstructed 
for the same sample of groups by \citet{wang12}. 
Briefly, \citet{wang12} reconstructed the velocity field by 
using galaxy groups/halos to reconstruct the matter density field 
and by using a quasi-linear perturbation model to estimate 
the velocity field from the reconstructed density field. 
Because the velocity field is coupled to the reconstruction 
of the density field through the correction of redshift distortion, 
an iteration procedure was applied to determine the final, 
converged velocity field. Tests with mock 
catalogs show that the errors in the reconstructed peculiar 
velocities have a symmetric distribution around zero, with dispersion
of about $90\,{\rm km\,s^{-1}}$. Our final sample contains 
all groups with $z\le 0.12$, within which groups with 
$M_{200}>10^{12.5}\,h^{-1}{\rm M_\odot}$ are complete \citep{yang07}.


\section[methods]{METHOD AND ANALYSIS}
\label{sec_methods}

\subsection{Extracting the kSZE signal}
\label{ssec_kSZ}

The CMB spectrum is distorted when CMB photons interact with 
free electrons that are moving collectively. In this kinetic 
Sunyaev-Zel'dovich effect (kSZE), temperature change
is characterized by a dimensionless parameter, 
\begin{eqnarray}\label{eq_k}
k\equiv \left(\frac{\Delta T}{T_{\rm CMB}}\right)_{\rm kSZ}
 = -\frac{\sigma_{\rm T}}{c} \int{n_{\rm e} (\bm{v}\cdot\hat{\bm r}) {\rm d}l}, 
\end{eqnarray}
where $\bm{v}$ is the velocity of bulk motion, and $\hat{\bm r}$ is 
the unit vector along a line-of-sight (LOS). Assuming that electrons are moving 
together with the galaxy system containing them, which is justified
by the fact that the correlation length of the peculiar 
velocity field is much larger than a halo size \citep[e.g.][]{hand12, HM15}, 
we have   
\begin{eqnarray}\label{eq_tau}
k = -\frac{v_r}{c}\tau, 
~~~~~~
\tau (R)=\sigma_{\rm T}\int{n_{\rm e} (\sqrt{R^2+l^2}){\rm d}l}
\end{eqnarray}
where $v_r$ is the CMB rest-frame peculiar velocity of the galaxy system 
along the LOS. In our analysis, we assume a fixed profile for $n_{\rm e}$. 
Specifically, we adopt an empirical $\beta$-profile, 
\begin{eqnarray}\label{eq_n_e}
n_{\rm e}(r)=n_{\rm e,0}[1+(r/r_{\rm c})^2]^{-3\beta/2},  
\end{eqnarray}
where $r_{\rm c}=r_{\rm vir}/c$ is the core radius of a group with 
concentration $c$, and $\beta=0.86$ is the best-fit value
obtained from South Pole Telescope (SPT) cluster profiles \citep{plagge10}. 
Note that $r_{\rm vir}$ and $c$ are determined by the halo mass and 
redshift of the galaxy group in question. In principle, the integrated 
signals extracted can depend on the assumed gas profile. However, 
because of the beam size of the {\it Planck} instrument, which is comparable 
to halo radius for a significant fraction of the groups in our sample, 
the impact of the specific choice for the spatial profile shape is 
significantly mitigated, as is discussed in \S\ref{ssec_test} and 
\S\ref{ssec_uncertainty}. 

 To extract the kSZE of a group from the {\it Planck} data, we apply a filter 
that takes into account the effect of the beam, 
\begin{eqnarray}\label{eq_MMF}
\hat{F}_I(\boldsymbol{k}) = \hat{\tau} (\boldsymbol{k})\, 
\hat{B}_I(\boldsymbol{k})\,, 
\end{eqnarray}
where $\hat{F}_I(\boldsymbol{k})$ is the Fourier transform of the 
filter for each of the three frequencies, `$I$', $\hat{\tau}(\boldsymbol{k})$ 
is the Fourier transform of $\tau(R)$, 
and $\hat{B}_I(\boldsymbol{k})$ is the Fourier transform of the 
Gaussian beam function that mimics the convolution of the 
{\it Planck} observation in the 
frequency band `$I$'. We use the filter, $\{F_i\}_I$, defined in equation 
(\ref{eq_MMF}) with redshift, virial radius and concentration appropriate 
for the group, `$i$', in question, at the frequency `$I$' 
to estimate the signal within the filter,     
\begin{equation}\label{eq_filter}
\{A_{D,i}\}_I = \int{\{F_i(\boldsymbol{\theta})\}_I
\{D_i(\boldsymbol{\theta})\}_I{\rm d}^2\theta}\,,  
\end{equation}
where $\boldsymbol{\theta}$ is the projected position relative to the 
group center in the sky, and $\{D_i(\boldsymbol{\theta})\}_I$ is 
the data map around the group. 

The values of $\{A_{D,i}\}_I$ obtained this way can be affected by projection 
effects by other groups along the same line of sight. Because of this, 
we do not use these values to represent the signal produced by individual 
groups and to obtain the average quantities for groups of a given mass. 
Instead, we construct model maps that take into account the projection 
effects to compare with data. As described above, the kSZE signal expected 
from a given group is determined by its peculiar velocity and its gas density 
profile, which is modeled by equation (\ref{eq_n_e}). In our analysis, we use 
the reconstructed peculiar velocities described in \S\ref{ssec_group}. 
The virial radius, $r_{\rm vir}$, and concentration, $c$, of a group 
are given by its halo mass, as described above. To model the total 
ionized mass associated with a halo, we assume that the 
amplitude of the profile, $n_{\rm e, 0}$, depends only on halo mass. 
We specify the halo mass dependence by the values of $n_{\rm e, 0}$
at $\log (M_{500}/{\rm M_\odot}) = 12.3$, $12.7$, $13.1$, $13.5$, $13.9$, 
and $14.3$, together with linear interpolations in 
$\log(n_{\rm e, 0})$ - $\log (M_{500})$ space to predict the profile 
amplitude for any given $M_{500}$. Thus, the model is completely specified by a 
set of six model parameters that give the profile amplitudes at the six 
values of $M_{500}$.  The numbers of groups in 
the six mass bins are listed in Table\,\ref{tab_K200}, together 
with the median values of $M_{200}$, defined in a way similar to 
$M_{500}$. For a given set of model parameters, 
denoted collectively by $\Theta$, we generate a theoretical map by convolving 
the center of each group with the 2-d profile appropriate for its halo mass, 
redshift and peculiar velocity. 
We convolve the resulting map with the beam function to mimick 
the beam effect of the observation. 
Note that the model maps include the 
projection effects generated by the superposition of the profiles of halos 
along a LOS, because they are constructed with the positions of all groups. 
Then, exactly the same method is used to extract signals from the model maps. 
We put the filter at the center of each of all 
the groups in the model map, again according to the halo mass 
and redshift of the group, to extract the corresponding model signal, 
\begin{equation}
\{A_{M,i}\}_I= \int{\{F_i(\boldsymbol{\theta})\}_I
\{M_i(\boldsymbol{\theta})\}_I{\rm d}^2\theta},
\end{equation}
where $\{M_i(\boldsymbol{\theta})\}_I$ stands for the model map at 
frequency `$I$' around group `$i$'. 

With these measurements, we compute a $\chi^2$-based likelihood function, 
\begin{equation}
{\cal L}(\Theta|A_D) \propto \exp(-\chi^2/2), 
\end{equation}
where 
\begin{equation}
\chi^2 = {\Big[{\boldsymbol A}_{D}-{\boldsymbol A}_{M}(\Theta)\Big]^T
{\rm {\bf Cov}}^{-1}
\Big[{\boldsymbol A}_{D}-{\boldsymbol A}_{M}(\Theta)\Big]}\,.
\end{equation} 
where ${\boldsymbol A}$'s are one-dimensional arrays with the elements being 
$\{A_{i}\}_I$ for all groups and in the three frequency bands, thus 
each containing $3N_{\rm grp, tot}$ elements, with $N_{\rm grp, tot}$
being the total number of groups. 
The covariance of the estimate between group `$i$' in band 
`$I$' and group `$j$' in band `$J$', 
\begin{eqnarray}
{\rm Cov}_{ij, IJ} = \Big\langle
\big(\{A_{D,i}\}_I-\{A_{M,i}\}_I\big) 
\big(\{A_{D,j}\}_J-\{A_{M,j}\}_J\big)
\Big\rangle , \nonumber \\  
\end{eqnarray}	
is calculated from $1,000$ sets of random shift and rotation of 
the filters relative to the CMB sky but with 
their relative positions fixed, by applying the filters 
to the maps as in equation (\ref{eq_filter}). 
This way, the values returned by the filters measure the 
total noise fluctuation while the spatial correlation of the noise 
among groups is retained. We use the average correlation of the values 
from these $1,000$ random sets to estimate the covariance matrix. 

Note that the covariance computed this way takes into account all noise 
components that are not correlated with the targets and are 
homogeneous over the sky, including the primary CMB fluctuation, 
instrument noise, dust emission, point sources that are not 
masked, and residual tSZE and kSZE from sources that are not 
correlated with our targets. The primary CMB 
fluctuation is treated as noise because the maps 
we use are not cleaned of it. It is very difficult to 
separate the primary CMB from kSZE because they have the 
same frequency dependence. Our noise estimate may 
also contain parts of the kSZE from the targets,  
because the targets cover a significant fraction of the sky 
and so some filters from the $1,000$ sets of random shift 
and rotation may end up in regions covered by the targets. 
The impact of this, however, must be negligible given that the 
kSZE is much weaker than other components of the noise, 
in particular, the primordial CMB. 
The error in the velocity reconstruction and 
the intrinsic variance of the kSZE at given mass can, 
can, in principle, also be included in the covariance matrix. 
However, the uncertainty expected from those errors combined 
is only a few percent of the primary CMB fluctuation, and 
can safely be ignored without affecting our results.
Uncertainties produced by the contamination that is correlated 
with the targets are not included in our covariance matrix.  
We will test the effect of such contamination in 
\S\ref{ssec_uncertainty}. 

To efficiently explore the parameter space, we make 
use of the MULTINEST method developed in \citet{feroz09}, which implements 
the nested sampling algorithm developed in \citet{skilling06}.  
The posterior distribution of the model parameters, i.e. the values 
of $n_{\rm e, 0}$ at the six halo masses, is used to make inferences 
from the data. 

The amplitude of the density profile obtained from the filters, 
together with the assumed spatial profile shape, can be 
used to estimate the total number of electrons within $R_{500}$ 
for a group of a given mass, 
\begin{eqnarray}
d_{\rm A}(z)^2K_{500}\equiv \sigma_{\rm T} \int_{R_{500}}{n_{\rm e}{\rm d}V}\,.
\end{eqnarray}
where $d_{\rm A}(z)$ is the angular diameter distance 
of the group in question. We also define the following quantity, 
analogous to the convention used in the tSZE studies, 
\begin{eqnarray}\label{eq_pl}
\tilde{K}_{500}\equiv K_{500} (d_{\rm A}(z) / 500 {\rm Mpc})^2\,.
\end{eqnarray}
We emphasize, however, that we are not measuring $\tilde{K}_{500}$ directly 
from the data, but using it to represent the amplitude of the density profile, 
$n_{\rm e}$. Once $\tilde{K}_{500}$ is obtained, we can also 
obtain corresponding quantities within other choices of
radius, such as $\tilde{K}_{200}$ within $R_{200}$.

\subsection{Testing with a mock observation}
\label{ssec_test}

\begin{figure*}
\includegraphics[width=1.\linewidth]{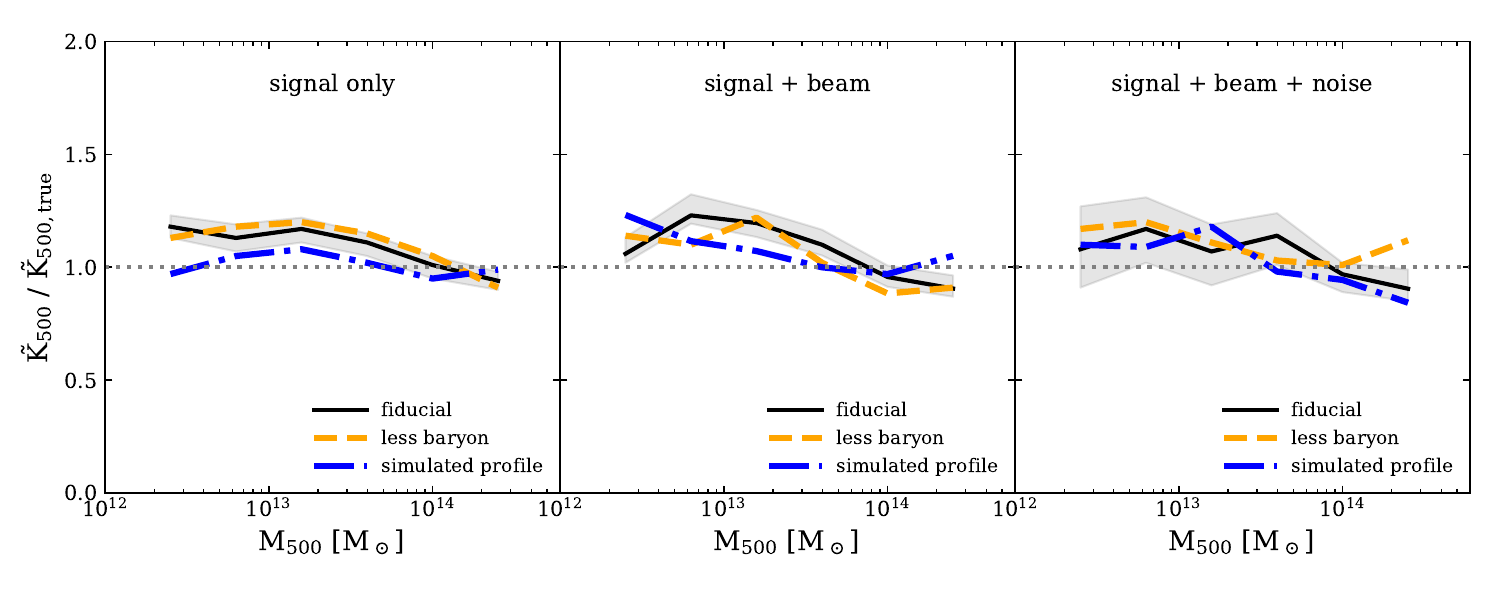}
\caption{The $\tilde{K}_{500}$ recovered by our method from the simulated map, 
with respect to the true $\tilde{K}_{500}$ directly from 
the simulation, for the `signal-only' map (left), the `noise-free' map (middle), and 
the realistic map (right) (see the text for details on how the maps are constructed). 
The black solid lines show 
the results for the fiducial case with the bands showing the $68\%$ ranges of the 
posterior distribution. The dashed lines show the results for a model where 
the baryon fraction equals the universal fraction for the most massive bin but 
decreases with halo mass by a power-law to be one third of the universal fraction 
at the lowest mass bin. 
The blue dot-dashed lines show the results when the profile directly measured 
from the simulation is used for the method. }
\label{fig_test}
\end{figure*}

In this section, we test our method described above by applying it to  
mock observation and by examining how our method recovers the true 
$\tilde{K}_{500}$. The mock {\it Planck} map is constructed from 
the Magneticum \citep{dolag16}, a set of cosmological gas simulations 
of various volumes and resolutions, run with an improved version of 
GADGET-3 \citep{springel05} and with WMAP7 cosmology 
\citep{komatsu11}. The specific run chosen for 
our analysis samples a box of $L=352\,h^{-1}{\rm Mpc}$ with 
$2\times (1584)^3$ particles, which results in a mass resolution 
of $m_{\rm DM}\sim 10^9 {\rm M_\odot}$ for the dark matter 
particles. Based on the simulation, we construct the kSZE light-cone maps 
using the SMAC code \citep{dolag05}, which cover the redshift range 
of $0.0173 <z< 0.194$ and $900\,{\rm deg}^2$ of the sky. We generate the kSZE 
based on the dark matter particles only, assuming that gas particles follow 
the dark matter distribution with an assumed ratio between the gas and 
total mass densities. We then generate the CMB temperature maps at the same 
three frequencies as used in our analysis, and add the primordial CMB 
anisotropy using the CMB power spectrum. Finally, we degrade the map with the 
{\it Planck} beam function, and add the instrument noise. 
To mimic the impact of the primordial CMB anisotropy and the instrument 
noise on our results based on a sample that is nine times larger 
in sky coverage than the mock sample considered here, we 
reduce their amplitudes by a factor of $\sqrt{9}$. We neglect 
other contaminating sources, such as dust emission, residual tSZE and kSZE, 
because they are negligible compared to the primary CMB fluctuations 
at the frequencies considered here. 

The three panels of Fig.\,\ref{fig_test} show the results of 
$\tilde{K}_{500}$ obtained by applying our method to the three 
stages of the constructed temperature maps, in comparison to the 
values obtained directly from the simulation. The left panel shows
the results for maps that do not contain beam smearing and 
noise, the middle panel contains only beam smearing, while the 
right panel contains both. The black solid line shows the result 
for all halos recovered by assuming the $\beta$ profile 
of equation (\ref{eq_n_e}), which we use as our fiducial case. 
The shaded bands show the $68\%$ range of the posterior distribution. 
The median $\tilde{K}_{500}$ recovered by our method is about 
$20\%$ higher for halos with $M_{500}\leq 10^{13.5}\,{\rm M_\odot}$ 
relative to the true value. The bias does not seem to depend 
strongly on whether or not beam effects and noise are included, 
and is within the statistical uncertainty expected from the 
observational data (see the right panel).  

 Since our method uses projected profiles (see equation \ref{eq_tau})
to extract kSZ signals, it has the ability to suppress 
the kSZE signals contributed by the gas that appears to be 
associated with the halos in question but is located outside 
the halos. However, it is likely that a fraction of the 
signals detected by our method actually comes from the gas 
that is not associated with the halos in our sample, given 
that the typical correlation length of the peculiar velocity 
field is large. This may explain why our method overestimates 
the kSZE for low-mass halos. To check the effect of such 
contamination further, we artificially reduce the gas 
fraction within halos, so as to increase the importance of background 
contamination. To do this, we assume that the gas fraction, 
adopted in constructing the mock maps, depends on halo 
mass through a simple power-law relation, so that the gas 
fraction changes from the universal fraction for the most 
massive bin to one third of the universal fraction for the 
least massive bin. The gas fraction associated with dark 
matter outside these halos is still assumed to be at the 
universal fraction. The results obtained by applying our method 
to the mock maps are shown by the yellow dashed line. As one can see, 
the bias in the recovered $\tilde{K}_{500}$ is similar to that in 
the fiducial case, indicating that the use of filtering profiles 
can effectively suppress contributions from gas outside halos. 

We have carried out another test in which the shape of density 
profiles used to model $n_{\rm e} (r)$ is directly measured from 
the simulation, instead of that given by the $\beta$-model.
Since the gas density in the mock maps is directly proportional 
to the dark matter density field, the profile shape used here 
is similar to that of dark matter halos. The results, shown by the 
blue dot-dashed lines, indicate that using the true profile shape
can reduce the bias when the beam effect and noise are 
negligible. This happens because the dark matter halo 
profiles are more concentrated than the $\beta$-model, so that 
the contribution from large projected distances is suppressed 
more effectively. However, as shown in the middle and right panels, 
the beam size and, particularly, the noise significantly reduce the 
sensitivity of the results to the choice of the profile shape. 

Based on the tests presented above, we conclude that our method 
may be biased toward a higher kSZE flux by up to $\sim 20\%$ 
for low-mass halos due to the contamination by the background 
gas, but the bias is not significant given the statistical 
uncertainty in the present data. Note that the tests here 
are aimed at assessing the contamination by the gas that is not associated 
with the groups/halos in our catalog. In a forthcoming paper 
\citep{Lim_etal2020a}, we use realistic mock CMB maps and 
group samples to check the impact of a number of other uncertainties. 
We find that the net bias produced by the uncertainties
in group identification and in halo mass assignment is not 
significant in the results obtained from current observational 
data. We also find that an error of $\sim 90\,{\rm km\ s^{-1}}$
in halo peculiar velocities, a value similar to that in the 
reconstruction used here, does not lead to any significant 
impact on our results. 

\begin{figure*}
\includegraphics[width=1.03\linewidth]{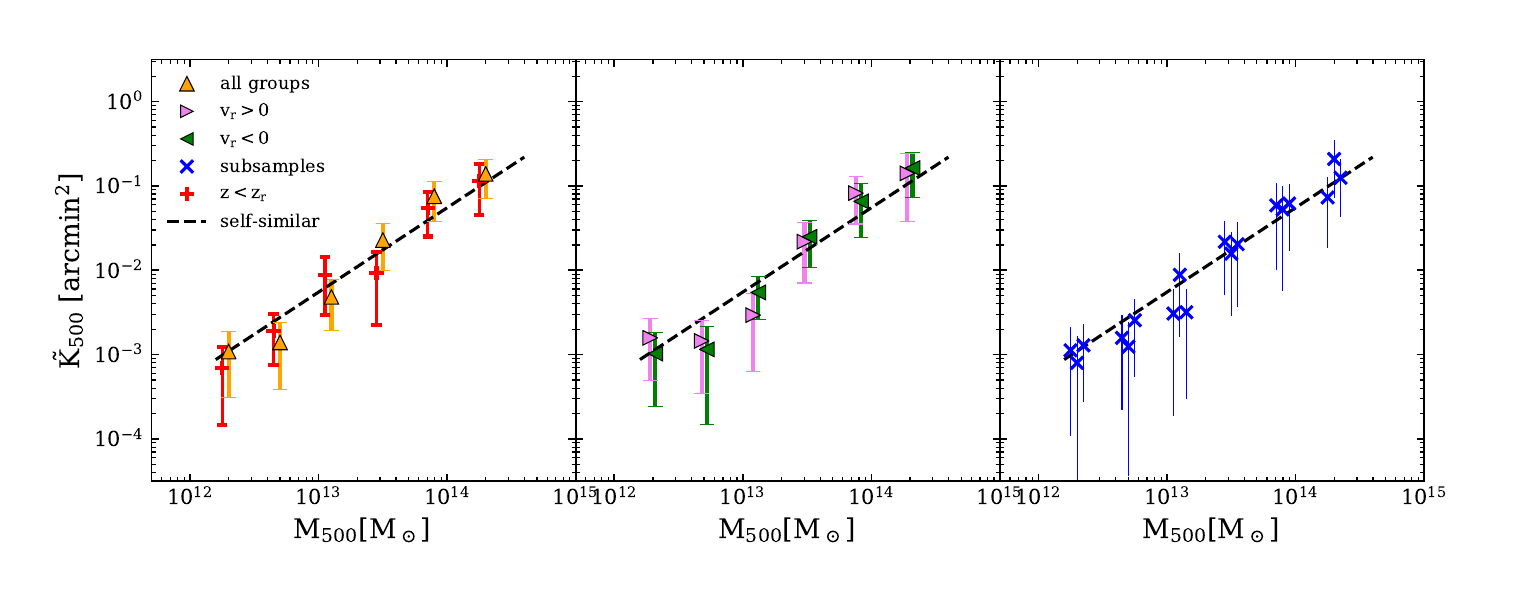}
\caption{The $\tilde{K}_{500}$-$M_{500}$ relations obtained from seven 
samples: 
all groups (yellow);
groups with $v_r>0$ (violet); 
groups with $v_r<0$ (green); 
groups in three parts of the sky (blue), 
and groups with $z<z_{\rm r}$ (red). 
The data points for some samples are shifted by $0.05$ dex 
horizontally for clarity. The dashed line shows the `self-similar' model, 
$N_{\rm e, 500} =[ (1+f_{\rm H}) / 2m_{\rm p}] \cdot f_{\rm B} M_{500}$, 
where $N_{\rm e,500}$ is the total number of electrons within $R_{500}$, 
$f_{\rm H}=0.76$ the hydrogen mass fraction, 
$m_{\rm p}$ the proton mass, and $f_{\rm B}=\Omega_{\rm B} / \Omega_m=0.16$ 
the cosmic baryon fraction. All the data points are based on the 
medians of the corresponding posterior distributions given 
by the MULTINEST sampler. The error bars indicate the $68$ percentile 
ranges of the corresponding posterior distributions. }
\label{fig_K500}
\end{figure*}

\section[results]{RESULTS}
\label{sec_results}

\subsection{The $\tilde{K}_{500}$-$M_{500}$ relation}
\label{ssec_K500}

The results obtained from the entire sample are shown as yellow 
triangles in Fig.\,\ref{fig_K500}. The data points represent the median 
values obtained from the posterior distribution, while the   
error bars are the $68$ percentile range. 
The dashed line in Fig.\,\ref{fig_K500} shows the `self-similar' 
model prediction in which the total number of electrons within 
$R_{500}$ is, 
\begin{eqnarray}
N_{\rm e, 500} =[ (1+f_{\rm H}) / 2m_{\rm p}] \cdot f_{\rm B} M_{500}
\end{eqnarray}
with $f_{\rm H}=0.76$ the hydrogen mass fraction, 
$m_{\rm p}$ the proton mass, and $f_{\rm B}=\Omega_{\rm B} / \Omega_m=0.16$ 
the universal baryon fraction. Our data points follow well the self-similar model, 
indicating that the total ionized gas fractions in halos of different masses 
are comparable to the universal baryon fraction 
(see \S\ref{ssec_gas} for the details).

\begin{figure}
\includegraphics[width=1.\linewidth]{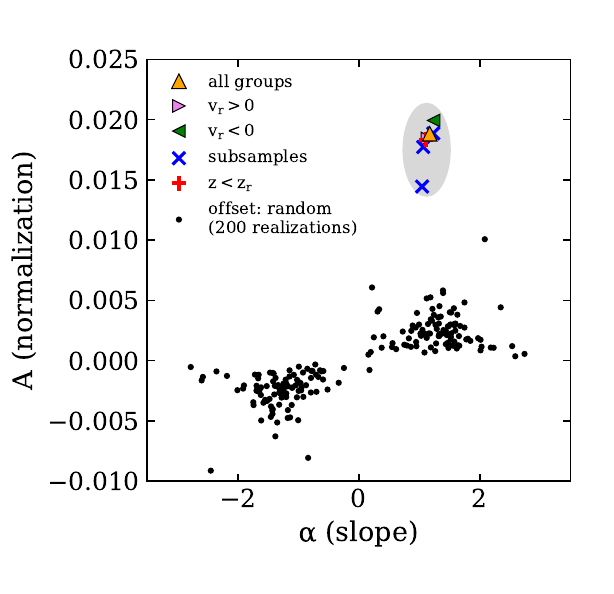}
\caption{The results obtained from the power-law model, 
$\tilde{K}_{500} = A \times (M_{500} / 10^{13.5}{\rm 
M_\odot})^\alpha$, for the seven observational samples
(coloured symbols), and for the $200$ random realizations, in which the 
total group sample is shifted and rotated by some random amounts (black dots). 
The shaded ellipse covers the area occupied by different observational samples, 
as indicated in the panel.}
\label{fig_A_alpha}
\end{figure}

Fig.\,\ref{fig_K500} shows that the $\tilde{K}_{500}$ - $M_{500}$ 
relation is approximately a power law. This motivates another way to 
extract the $\tilde{K}_{500}$ associated with groups. Here we assume that     
\begin{eqnarray}\label{eq_pl}
\tilde{K}_{500} = 
A \times (M_{500} / 10^{13.5}{\rm M_\odot})^\alpha
\end{eqnarray}
to generate the model maps, and use the same method of filtering 
and minimization of the $\chi^2$ to constrain $\alpha$ and $A$. 
To fit this relation, we neglect potential intrinsic scatter
of the relation and uncertainties in halo 
identifications, based on the test results described 
in \S\ref{ssec_test}. 
The result for the entire sample is shown in Fig.\,\ref{fig_A_alpha}, 
with a yellow triangle. 
As comparison, each of the small dots shows the result of a random sample. 
Each random sample is constructed by shifting and rotating the group 
sample by some random amounts relative to the {\it Planck} maps before 
applying the filter. In this case, the relative 
positions of individual groups and their spatial clustering are 
preserved in the random samples, but the cross correlation between 
the groups and the kSZE signals is destroyed.  
The distribution of the $200$ random realizations is around 
$(A,\alpha)= (0,0)$, as expected from a zeroed mean background. The symmetry 
relative to $(0,0)$ is because the peculiar velocities have a roughly symmetric 
distribution around zero. The dipolar pattern of the random samples 
in $(A, \alpha)$ space, that positive (negative) values of $A$ tend 
to correspond to positive (negative) values of $\alpha$, indicates the 
presence of residual background fluctuation on scales 
larger than individual groups. In this case, the flux 
associated with a group in a random sample due to background fluctuation 
is proportional to the angular size of the group, with approximately the same 
probability to be positive or negative. Thus, the distribution of 
random groups in the $\tilde{K}_{500}$-halo mass plane is expected to have a 
wedge-like pattern symmetric with respect to the halo-mass axis. 
Since we have more lower-mass systems, the averages of the 
$\tilde{K}_{500}$ in the lower-mass bins are closer to zero and have lower random 
fluctuations. This has the effect of increasing the opening 
angle of the wedge within which the average $\tilde{K}_{500}$ versus halo 
mass relations from different random samples are confined.  
The assembly of straight lines, each covering the whole halo mass 
range in such a wedge, thus tend to show the dipolar pattern seen.  
 
We have also used random samples in which each group 
in the observational sample is assigned a random position in the {\it Planck}
sky, and found very similar results. The results indicate that the detection of the 
$\tilde{K}_{500}$ is very significant relative to the random samples, as 
we will quantify below after the uncertainties in the measurements are tested.   
 
\subsection{Tests of uncertainties}
\label{ssec_uncertainty}

 A number of contaminating effects can affect our measurements. 
Here we present the analyses we have carried out
to test the reliability of our results against the 
contamination. 
One source of uncertainty comes from the errors in the peculiar 
velocities of groups adopted in our analysis. As shown in detail 
by \citet{wang12} using realistic mock samples, the errors in the 
reconstructed peculiar velocities have a symmetric distribution 
around zero. Since the kSZE of a group is directly proportional to 
its peculiar velocity, the uncertainty in the peculiar velocities 
is not expected to produce any bias in our results, but 
will contribute to noise in the measurements. 


 Another source of contamination is from fluctuations 
in the background and foreground, such as the primordial 
CMB, signals produced by background sources that are not 
included in our sample, and residual Galactic foreground. 
If the background/foreground fluctuations are not 
correlated with the groups in our sample, then the 
contamination is not expected to lead to any bias in our results, 
but can increase the noise in our measurements. 
To test this, we divide the total sample into three sub-samples,  
each containing groups in a $\sim 1/3$ portion of the sky coverage 
according to their Galactic latitudes, and repeat the 
procedures to obtain the $\tilde{K}_{500}$ for groups 
in each of the three sub-samples. The results of the three sub-samples, 
shown as the blue crosses in Fig.\,\ref{fig_K500}, are similar for 
all the mass bins. In addition, as shown in our test using random samples 
(Fig.\,\ref{fig_A_alpha}), any residual background/foreground 
fluctuations are well below the signals we detect. These tests show 
that this type of contamination does not have a significant impact on 
our results. 

 Yet another source of contamination comes from the emissions of the groups
in the observational wavebands, such as radio and infrared emissions 
and the tSZE. Although we have attempted to subtract the tSZE from the 
observational data, some residual may still exist. One unique property of 
the kSZE is that two similar groups with opposite peculiar velocities produce 
temperature fluctuations with opposite signs, in contrast to the
contaminating emissions mentioned above, which should be independent 
of the sign of the peculiar velocity. To check that the 
signals we detect are indeed produced by the kSZE, we divide the 
entire sample into two sub-samples, one with $v_r>0$ and the other with  
$v_r<0$, and tune the model parameters 
independently for the two sub-samples to achieve the best match to the data. 
The results obtained for these two sub-samples are shown in Fig.\,\ref{fig_K500}
as the violet and green triangles, respectively. 
The two give consistent $\tilde{K}_{500}$ in all the mass bins.
The fact that the sub-samples of opposite peculiar velocities give similar 
$\tilde{K}_{500}$ - $M_{500}$ relations [i.e. opposite signals in $k$ defined in 
equation (\ref{eq_k})] suggests that the contamination by emission sources 
does not change our results significantly, and that the signals we detect are 
indeed kSZE.

\begin{table*}
 \renewcommand{\arraystretch}{1.6} 
 \centering
 \begin{minipage}{116mm}
  \caption{The $\tilde{K}_{200}$-$M_{200}$ relation.}
  \begin{tabular}{ccccccc}
\hline
Bin no. & 1 & 2 & 3 & 4 & 5 & 6 \\
\hline
\hline
$\log M_{200}\,[{\rm M_\odot}]$ & 12.4 & 12.8 & 13.2 & 13.6 & 14.0 & 14.4 \\
number of systems & $23,997$ & $9,795$ & $3,780$ & $1,287$ 
& $346$ & $58$ \\
$\tilde{K}_{200}\,[10^{-2}{\rm arcmin}^2]$\textsuperscript{\footnote{The average 
of the median values of the posterior distribution from the seven samples used for our test of systematic effects.}} & 0.19 & 0.29 & 0.95 & 3.5 & 12 & 25 \\
$\sigma (\tilde{K}_{200})\,[10^{-2}{\rm arcmin}^2]$\textsuperscript{\footnote{The variance of the median values 
of the posterior distribution among the seven samples. }} 
& 0.059 & 0.096 & 0.51 & 1.1 & 2.2 & 8.4 \\
$\delta (\tilde{K}_{200})\,[10^{-2}{\rm arcmin}^2]$\textsuperscript{\footnote{The 
variance divided by $\sqrt{6}$ among the six independent sub-samples, each containing 
groups in a $\sim 1/6$ portion of the sky, 
to represent the statistical uncertainty in the estimates.}} & 0.063 & 0.064 & 0.49 & 
0.95 & 4.7 & 6.0 \\ 
$z_{\rm r}$\textsuperscript{\footnote{The redshift where the median $\theta_{200}$ of halos equals the 
{\it Planck} beam size of $10\arcmin$.}} & 0.029 & 0.041 & 0.058 & $0.081$ & $0.11$ & $>0.12$ \\
cov$(i,1)$\textsuperscript{\footnote{cov$(i,j)$ is the covariance of 
$\tilde{K}_{200}$ between $i$-th and $j$-th bins, calculated 
for the sample of all groups, according to the bin number assignment in the first column of the table. The values are in units of $10^{-4}{\rm arcmin}^4$. }} 
           & $(0.058)^2$ & & & & & \\
cov$(i,2)$ & -0.00080 & $(0.091)^2$ & & & & \\
cov$(i,3)$ & -0.0027 & -0.0027 & $(0.35)^2$ & & & \\
cov$(i,4)$ & -0.016 & -0.018 & -0.056 & $(1.5)^2$ & & \\
cov$(i,5)$ & -0.037 & -0.043 & -0.14 & -0.82 & $(5.1)^2$ & \\
cov$(i,6)$ & -0.071 & -0.078 & -0.23 & -1.4 & -3.3 & $(4.5)^2$ \\
\hline
\\
\vspace{-8mm}
\end{tabular}
\label{tab_K200}
\end{minipage}
\vspace{2mm}
\end{table*}

As detailed in \S\ref{ssec_test}, the large beam size of the {\it Planck} 
may contaminate the results with the kSZE signals from the gas outside halos. 
To test the contamination, we divide all halos into the `resolved' ones 
and the `unresolved' ones. We define a redshift, $z_r$, at which the median 
$\theta_{200}$ of the halos in a given mass bin is equal to the {\it Planck} 
beam size, $10\arcmin$. The values of $z_r$ for individual mass bins are listed in 
Table\,\ref{tab_K200}. For both the $z<z_r$ sub-sample (`resolved') 
and the $z>z_r$ sub-sample (`unresolved'), we assume two independent 
sets of parameters and jointly constrain them. 
The results so obtained for the `resolved' sub-samples are shown 
as the red crosses in Fig.\,\ref{fig_K500}. 
Clearly, the results obtained from the `resolved' sample 
are in good agreement with those obtained from the total sample within 
the uncertainty. 

Finally, our filter uses equation (\ref{eq_n_e}) to model the gas profile, while the 
true gas profiles may be different. To examine how our results are 
affected by the assumed profile, we have made tests by increasing the values
of $r_c$ by a constant factor, $\mu$. We found that our results do not change 
significantly as long as $\mu < 1.5$. When $\mu$ becomes larger than $1.5$, 
the values of $\tilde{K}_{500}$ obtained start to change noticeably 
for low-mass groups. The test shows that our results are not very sensitive to 
the gas density profile, as may be expected from the low angular 
resolution of the {\it Planck} data. However, this also indicates that 
the current data is not able to provide significant constraints on the 
details of the gas profiles around halos. A similar conclusion 
is reached in our forthcoming paper \citep{Lim_etal2020a}, 
where we test a range of profiles predicted from hydrodynamic 
simulations, using detailed mock CMB maps constructed to mimic the
{\it Planck} data. 

To summarize, all the seven samples we have analyzed above, 
namely all groups, groups with $v_r>0$, 
groups with $v_r<0$, groups in three parts of the sky, 
and groups with $z<z_{\rm r}$, give consistent 
results (as summarized in Fig.\,\ref{fig_K500}), 
demonstrating that our detection of the kSZE is reliable. 
For reference we list, in Table~\ref{tab_K200}, the averages 
and variances of ${\tilde K}_{200}$ for individual halo 
mass bins obtained from these seven samples. Since  
these samples are not independent, the variances  
listed can only provide a measure of systematic 
effects we have tested, but cannot be used to represent 
the statistical uncertainties in the estimates of the 
averages of ${\tilde K}_{200}$. To get some insight into 
the statistical uncertainties in our estimates, 
we divide the groups into 
six independent sub-samples, each containing groups in a 
$\sim 1/6$ portion of the sky coverage, and obtain 
${\tilde K}_{200}$ versus halo mass for each of the 
six sub-samples. The uncertainties in the estimates  
of ${\tilde K}_{200}$ obtained from these six independent 
sub-samples are also listed in Table\,\ref{tab_K200}. 
Table\,\ref{tab_K200} 
also lists the covariance matrix of ${\tilde K}_{200}$ among 
the six mass bins for the sample of all groups, 
obtained by drawing from the posterior distribution and then by 
calculating,  
\begin{eqnarray}\label{eq_cov}
{\rm {\bf cov}}\,=\,\left\langle\left({\tilde {\boldsymbol K}}_{200}-\langle{\tilde 
{\boldsymbol K}}_{200}\rangle\right)\left({\tilde {\boldsymbol K}}_{200}-
\langle{\tilde {\boldsymbol K}}_{200}\rangle\right)^{\rm T}\right\rangle
\end{eqnarray}
where `$\langle\cdot\cdot\cdot\rangle$' denotes the mean. As expected 
from the nature of our simultaneous constraining of the model parameters, 
the off-diagonal elements 
are all negative. To estimate the joint signal-to-noise from the six mass bins, 
we use the covariance matrix to define a multivariate normal distribution 
function, and calculate the cumulative probability of a null detection. 
The probability of our measurements thus calculated corresponds to  
a significance level of about $6\sigma$. 

The results with the power-law model [see equation \ref{eq_pl}] 
for all the seven samples we have analyzed above are 
shown as the colour points in  Fig.\,\ref{fig_A_alpha}. The data points 
are clustered in a region $(A,\alpha)\in (0.018\pm 0.0039, 1.1\pm 0.40)$.  
The dispersion among the seven samples is comparable to that of
$A$ at a given $\alpha$ obtained from the 200 random samples, 
indicating that the errors in the estimates are dominated by 
fluctuations in the background and foreground.
With this dispersion to represent the uncertainty in the 
results, our detection is at a level of more than $5\sigma$, 
consistent with the estimate from the binned data given above.

\subsection{The gas fraction and temperature}
\label{ssec_gas}

\begin{figure}
\includegraphics[width=1.\linewidth]{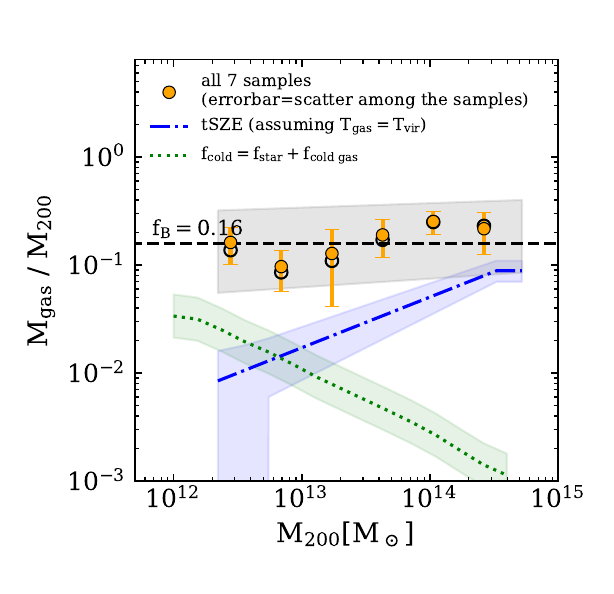}
\caption{The ratio between gas mass and halo mass within $R_{200}$ 
as a function of halo mass inferred from the observed 
$\tilde{K}_{500}$ - $M_{500}$ relations. Data points and error bars 
are the averages of, and the dispersion among, all the seven samples, 
respectively. 
The shaded band is based on the ellipse shown in Fig.\,\ref{fig_A_alpha}. 
The black unfilled circles show the results corrected for the bias 
according to the black curve in the left panel of Fig.\,\ref{fig_test}. 
The dashed line shows the universal baryon fraction of $f_{\rm B}=0.16$. 
The dot-dashed line represents the gas mass fraction
inferred from the tSZE by \citet{lim18a} assuming the gas 
to be at the virial temperature, with the shaded band indicating
the typical uncertainties in the data. The dotted line shows the 
mass fraction in stars \citep{lim17a} and cold gas \citep{popping14}, 
with the shaded band indicating the typical uncertainties in the data.}
\label{fig_Mgas}
\end{figure}

\begin{figure}
\includegraphics[width=1.\linewidth]{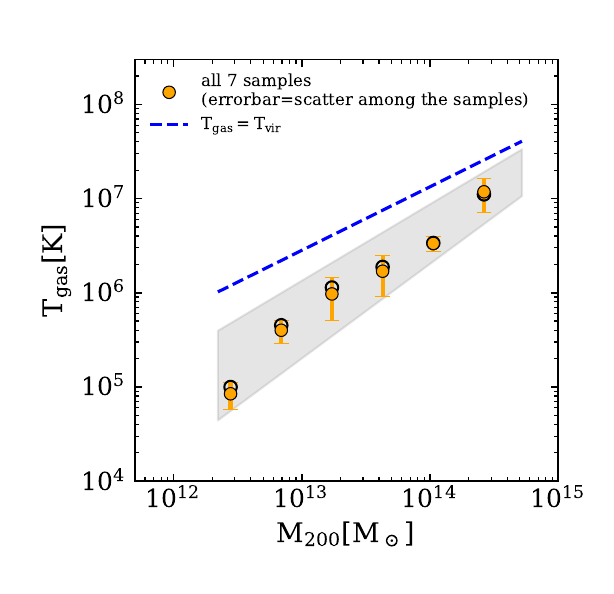}
\caption{The effective gas temperature obtained by dividing the electron thermal energy 
obtained from the tSZE measurement by the total number of electrons 
obtained from the kSZE. Data points and error bars 
show the averages of, and the dispersion among, all the seven samples,
respectively. 
The black unfilled circles show the results corrected for the bias 
according to the black curve in the left panel of Fig.\,\ref{fig_test}. 
The shaded band is based on the ellipse shown in Fig.\,\ref{fig_A_alpha}. 
The dashed line shows the virial temperature as a function of halo mass.}
\label{fig_Tgas}
\end{figure}

Fig.\,\ref{fig_Mgas} shows the gas fraction within $R_{200}$ obtained 
from the best-fit parameters and the assumed gas profile. 
The data points are the averages of and the dispersion among the seven samples, 
while the shaded band contains the predictions of the power laws enclosed 
by the ellipse in Fig.\,\ref{fig_A_alpha}. We also show the results corrected 
for the LOS contamination by the empty circles according to the black 
curve in the left panel of 
Fig.\,\ref{fig_test}. The inferred gas fraction is consistent with the 
universal baryon fraction, given the uncertainty of the data, and is 
much higher than the baryon fraction in stars and cold gas 
(shown as the dotted line). 

Recently, \citet{lim18a} measured the tSZE produced by 
galaxy groups using the nearly all-sky group catalog of \citet{lim17b}, 
which was constructed by applying the halo-based group finder to four large redshift 
surveys. The gas fraction inferred from the tSZE assuming
the virial temperature,  $T_{\rm vir}=\mu m_{\rm p}GM_{200} / 2k_{\rm B}R_{200}$,
with $\mu=0.59$ the mean molecular weight, is shown in Fig.\,\ref{fig_Mgas}. 
This fraction is much lower than that given by our kSZE data
except for the most massive groups, indicating that   
the average temperature of the gas responsible for the kSZE 
in lower-mass groups is much lower than the virial temperature. 
The effective temperature, estimated by combining the gas mass obtained 
from the kSZE and the thermal energy content given by the tSZE, is shown as a 
function of halo mass in Fig.\,\ref{fig_Tgas}. The derived 
effective temperature is about $10^5$-$10^6 {\rm K}$ for halos with 
$M_{200}\leq 10^{13.5} {\rm M_\odot}$, and much lower than the corresponding 
virial temperatures. This relatively low temperature has its origin 
from the relatively low thermal pressure measured in \citet{lim18a}, and we 
refer the reader to that paper for a detailed discussion about the 
comparison with other tSZE measurements. 

The temperature obtained here can be compared with that obtained from 
X-ray observations. As shown in \citet{pratt07}, the gas 
temperature in clusters of galaxies appears to decline in 
the outer parts. Since the effective temperature inferred 
from the SZE is sensitive to the low-density gas in the outer 
parts, this may partly explain the lower temperature found here 
than that inferred from X-ray observations. 
In addition, as shown in \citet{wang14}, for a given halo mass, 
the scatter in the X-ray luminosity is very large; at 
$M_{200}\sim 10^{13.5-14} {\rm M_\odot}$, the scatter 
is more than one order of magnitude (see their figure 7). 
As our SZE measurement is the average over all systems of a given halo 
mass, the thermal content inferred from it is expected 
to be lower than that obtained from X-ray selected samples
\citep[e.g.][]{pcx}. 

\section[summary]{SUMMARY AND CONCLUSION}
\label{sec_sum}

We have examined the kinetic Sunyaev-Zel'dovich effect (kSZE) 
from gas in dark matter halos associated with galaxy groups 
as a function of halo mass 
down to $\log (M_{500}/{\rm M_\odot}) \sim 12.3$. 
Our analysis uses the stacking of about $40,000$ galaxy 
groups to extract the kSZE from the {\it Planck} temperature maps in 
three different frequency bands, and employs a filter to 
take into account the beam effect. 
The filters are applied simultaneously for individual groups 
so as to minimize projection effects of halos along the same LOS. Accurate  
reconstructed peculiar velocities of the groups are used 
so that we can convert reliably the observed kSZE to the 
amounts of ionized gas associated with galaxy groups. 
A number of tests are made to examine the uncertainties in 
our results, from residual background/foreground fluctuations, 
from contamination by the tSZE and emissions from galaxy groups, 
from the large beam size of {\it Planck} observation,
from contamination by large-scale coherent motion,
and from the gas density profile adopted. 
We found that our results are robust against these potential sources 
of uncertainties for the current data. In a forthcoming paper, 
we will check in detail the significance of these uncertainties 
by applying the methods in our analysis and 
in the literature to the SZE light-cone maps generated from 
hydrodynamic simulations. 

 The strength of the kSZE as a function of halo mass is 
found to be consistent with the `self-similar' model, 
in which the baryon fraction is independent  of halo mass, 
suggesting that the `missing baryons' on halo scales 
are found by the kSZE. Combined with the tSZE measured for galaxy groups
by \citet{lim18a}, our results indicate that the gas temperatures in low-mass 
halos are much lower than the corresponding halo virial 
temperatures. This suggests that it is the low temperature 
of the gas, not the total amount of baryons, that is 
responsible for the low thermal energy contents in 
low-mass halos found in tSZE and X-ray observations.  
Our results, therefore, provide direct support to the 
hypothesis that the missing baryons in galaxy groups 
are contained in the WHIM with temperatures 
$\sim 10^5$ - $10^6$ Kelvin. 

Our results also demonstrate the potential of using the SZE to 
study both circum-galactic media (CGM) and galaxy formation 
processes that produce them. Such studies have advantages over 
absorption line studies, in that they are not 
limited to a small number of lines of sight, 
and that gas metallicity and ionization states are 
not needed to obtain the total gas mass. In the future, when 
high-resolution SZE data are available, the same analysis 
as carried out here can be used to constrain not only the 
total amount of ionized gas associated with galaxy groups 
(dark matter halos), but also to investigate the density and 
temperature profiles of the gas around them. One may also use 
galaxy groups with different star formation and/or AGN 
activities to study how the ionized gas distribution is 
affected by these activities. Clearly, the synergy 
between the SZE and observations of galaxy systems in other 
wavebands should be exploited in the future to provide detailed 
information both about the WHIM and about the galaxy formation 
processes that produce them.  

\section*{ACKNOWLEDGEMENTS}

We thank the Planck collaboration for making the full-sky maps public, 
and thank Klaus Dolag for providing the SZE light-cone maps. 
We thank Eiichiro Komatsu, Colin Hill, Neal Katz, Volker Springel, 
Mark Halpern, Gary Hinshaw, Ludo Van Waerbeke, Douglas Scott, and 
the referee for helpful discussions and comments, which greatly 
improved the paper. This work is supported by the 
China 973 Program (No. 2015CB857002) and the National Science 
Foundation of China (grant Nos. 11233005, 11621303, 11522324, 
11421303, 11503065, 11673015, 11733004). SL acknowledges a partial 
support by a CITA National Fellowship.

\end{document}